\documentclass{article}
\usepackage{color}
\usepackage{graphicx}
\usepackage{dcolumn}
\usepackage{bm}
\usepackage{physics}
\usepackage{blindtext}
\usepackage{float}
\usepackage{graphicx,wrapfig}
\newcommand*\diff{\mathop{}\!\mathrm{d}}

\newcommand*\Laplace{\mathop{}\!\mathbin\bigtriangleup}
\newcommand*\Prob{P\left(r,t | r_0\right)}
\newcommand*\Probs{p\left(r,t | r_0\right)}
\newtheorem{definition}{Definition}
 \usepackage{geometry}
 \usepackage{authblk}
\begin{document}

\title{Analytical Derivation of the Impulse Response for the Bounded 2-D Diffusion Channel}

\author[1,2]{Fatih Dinc}
\author[1]{Bayram Cevdet Akdeniz}
\author[3]{Ecda Erol}
\author[4]{Dilara G\"okay}
\author[1]{Ezgi Tekg\"ul}
\author[1]{Ali Emre Pusane}
\author[4]{Tuna Tugcu}
\affil[1]{Electrical \& Electronics Engineering Department, Bo\u{g}azi\c{c}i University, Istanbul, 34342, Turkey}
\affil[2]{Perimeter Institute for Theoretical Physics, Waterloo, Ontario, N2L 2Y5, Canada}
\affil[3]{Mechanical Engineering Department, Bo\u{g}azi\c{c}i University, Istanbul, 34342, Turkey}
\affil[4]{Computer Engineering Department, Bo\u{g}azi\c{c}i University, Istanbul, 34342, Turkey}

\date{Version: \today} 
\maketitle

\begin{abstract}
This paper focuses on the  derivation of the distribution of diffused particles absorbed by an agent in a bounded environment. In particular, we analogously consider to derive the impulse response of a molecular communication channel in 2-D and 3-D environment. In 2-D, the channel involves a point transmitter that releases molecules to a circular absorbing  receiver that absorbs incoming molecules in an environment surrounded by a circular reflecting boundary. Considering this setup, the joint distribution of the molecules on the circular absorbing receiver with respect to time and angle is derived. Using this distribution, the channel characteristics are examined. Furthermore, we also extend this channel model to 3-D using a cylindrical receiver and investigate the channel properties. We also propose how to obtain an analytical solution for the unbounded 2-D channel from our derived solutions, as no analytical derivation for this channel is present in the literature.
\end{abstract}


%

\section{Introduction}

Molecular communication (MC) has gained much attention recently as a promising method for communication among~nanodevices. An agent of such communication is the diffusion of molecules in biological environments, where the messenger molecules are used to mediate signals between transmitters and receivers. Due to the biocompatability of proposed nanodevices, medical applications constitute a promising application field. Therefore, examining the response of molecular communication channels is an important task to determine communication characteristics and possible communication scenarios. 

The receivers in molecular communication channels can be either absorbing that consume the incoming molecules, or observing that track the number of molecules inside a volume without absorbing them.
In the literature, impulse response for both types of channel models have been derived. In general, these channels can be categorized into two groups according to their environments. While one group of channels is placed in a free unbounded environment, the other group is placed in a bounded (and usually tubular) environment. For the first group, in \cite{eckford2007nanoscale}, the impulse response for a 1D channel is derived, while in \cite{yilmaz2014three} the 3D channel's impulse response is examined for a point transmitter and a spherical absorbing receiver under angular symmetry assumption. On the other hand, impulse response in a 2-D unbounded medium for a point transmitter and a circular absorbing receiver has not been derived, except for some special cases presented scenarios in \cite{akdeniz20172,dy2008first}. Not only the channels with point transmitters, but also the ones with spherical transmitters are considered in the 3-D medium in \cite{genc2018reception} and \cite{noel2016channel}.


As stated in \cite{gine2009molecular}, vessel-like channels  have beneficial effects for long-range molecular communication by preserving released molecules in a bounded range. Therefore, they have higher power efficiency, which is one of the reasons why many biological systems evolved in this direction. Since the molecules are not dispersed too much compared to the case of unbounded environments and due to their possible practical use in health applications, bounded and particularly vessel-like channels gained much attention in the literature. In \cite{turan2018channel}, 1-D and 3-D hitting location distribution of messenger molecules to a planar receiver is examined when there is no flow in the vessel-like environment. In \cite{wicke2017modeling} and \cite{dinc2018general}, the impulse response of a 3-D vessel-like channel is obtained for a spherical observer receiver when there is a laminar flow in the environment. In \cite{bicen2013system}, the flow models of microfluidic channels with different cross-section areas are presented, and the impulse response is derived by solving the 1-D diffusion-advection equation, which is only valid for some specific cases. Besides the channel impulse response, the capacity of the single-input single-output molecular communication channels with flow and drift is derived in \cite{sun2017channel}. In \cite{new}, an angle dependent approach is taken to consider a diffusion-based molecular communication system in a biological cylindrical environment.

In the nature, it is quite common to encounter diffusion processes that are bounded by membranes. One such example is the transmission of messengers inside a spherical cell bounded by the cell membrane, which can be modelled by a diffusion channel consisting of an absorbing spherical receiver and a reflecting spherical boundary \cite{dinc2018impulse}. In general, the spherical model is not enough to describe the diffusion processes inside the cell. In some cases, a cylindrical cell model can be more accurate. There are similar structures in the living organisms, like oval cells in the liver or simple columnar epithelium. In this paper, we derive the impulse response of a concentric cylindrical diffusion channel that involves a point transmitter and a cylindrical absorbing receiver to describe the diffusion inside such systems. Due to the symmetry of the system, we show that it can be reduced to a channel with a point transmitter and a circular absorbing receiver in a 2-D environment bounded by a reflecting circle. Therefore, our derived formula finds the impulse response of not only a microfluidic channel with an absorbing cylindrical receiver, but also of a channel in a 2-D environment bounded by a reflective circle, a point transmitter, and a circular receiver. Furthermore, we derive the generalized angle dependent impulse response for an annular channel, where the receiver only counts certain particles, which are absorbed inside the angle range $[-\theta_f,\theta_f]$. In summary, this paper deals with obtaining analytical expressions for:

\begin{itemize}
    \item  the impulse response  of the microfluidic channel with a point transmitter and a cylindrical receiver,
     \item  the impulse response of a 2-D environment with a reflective circular boundary, a point transmitter, and a circular absorbing receiver,
     \item the angle dependent characteristics of the channel impulse response.
\end{itemize}

\begin{figure*}
\centering
\includegraphics[width=7cm]{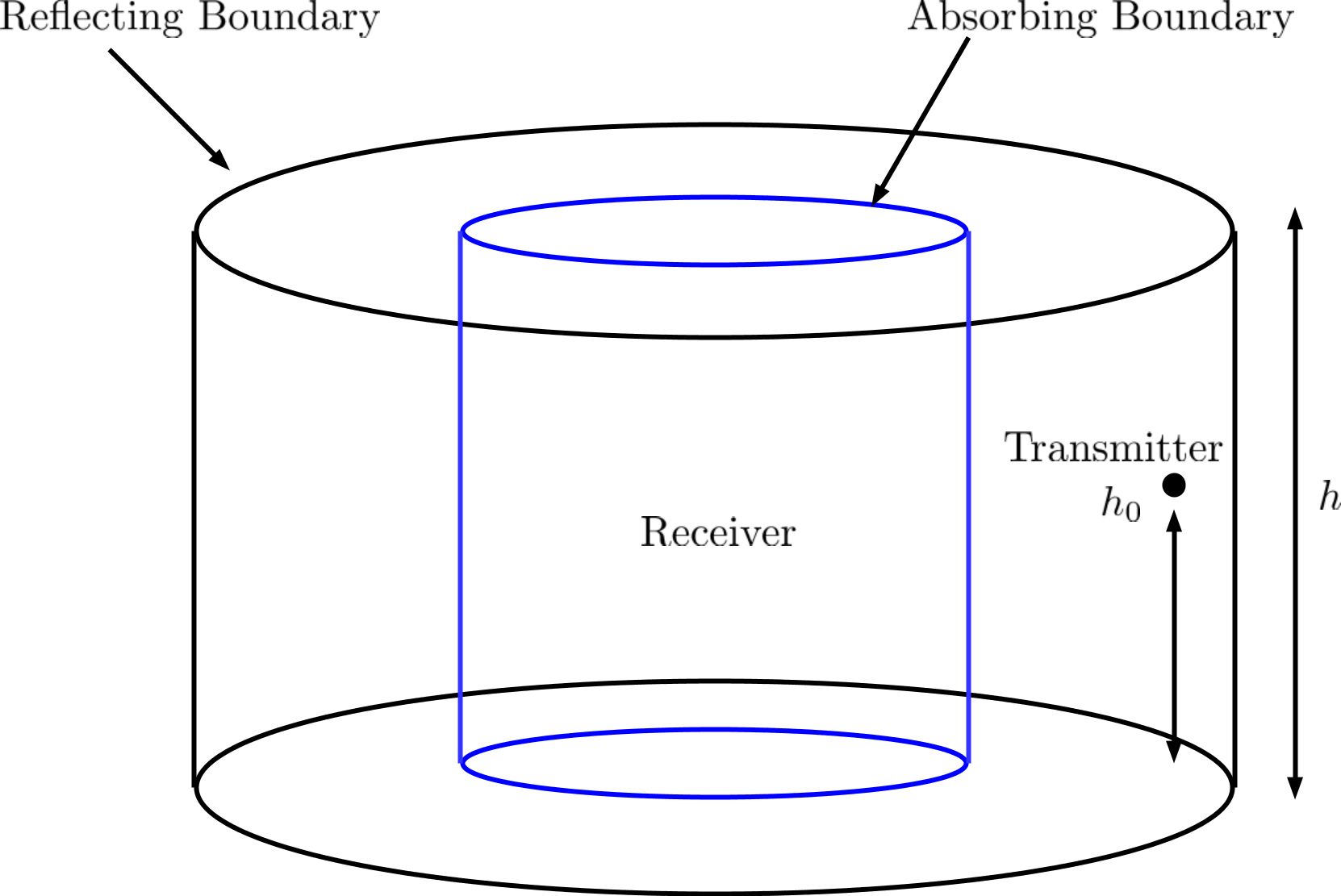} 
\includegraphics[width=4.3cm]{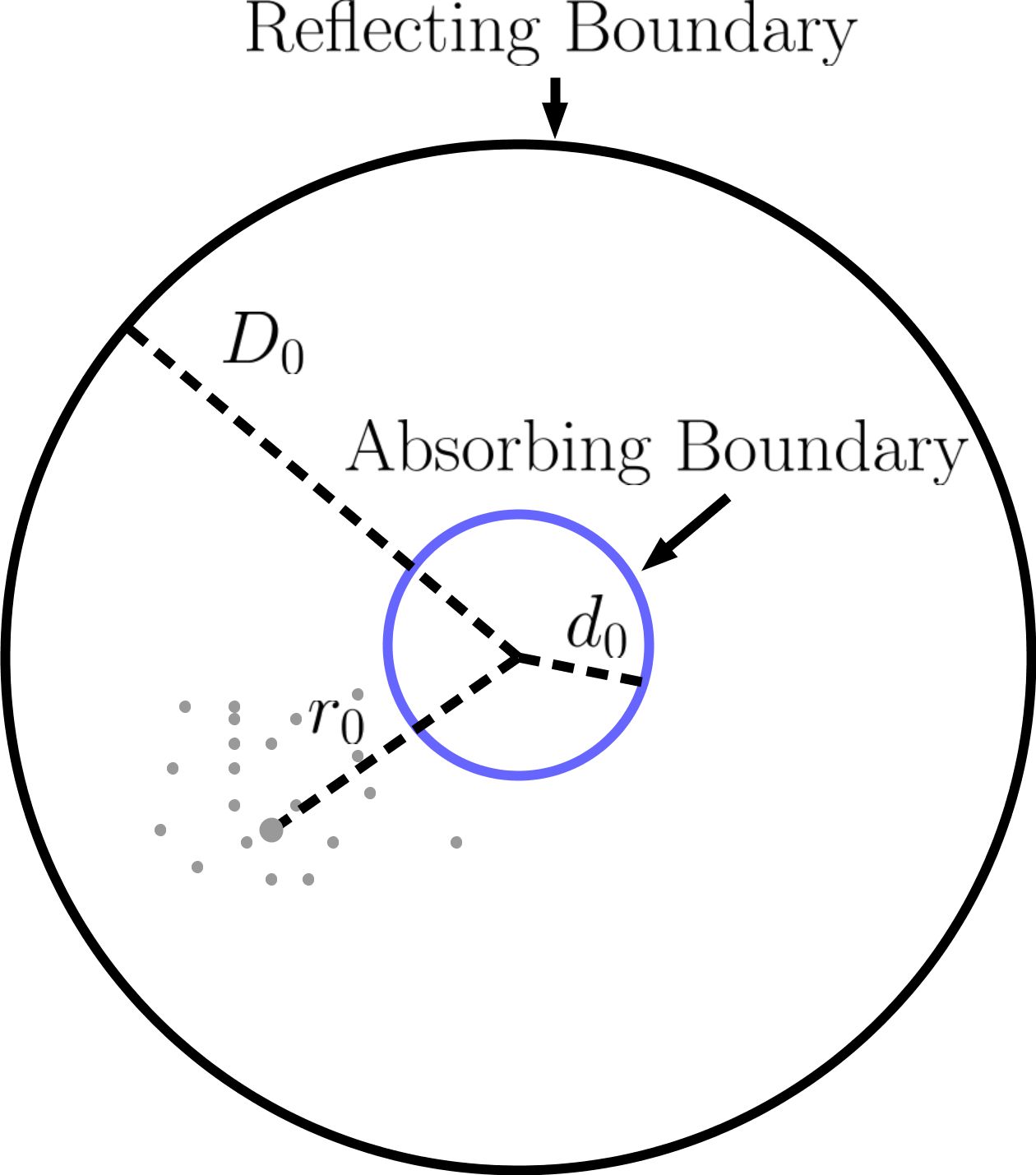} 
\includegraphics[width=4cm]{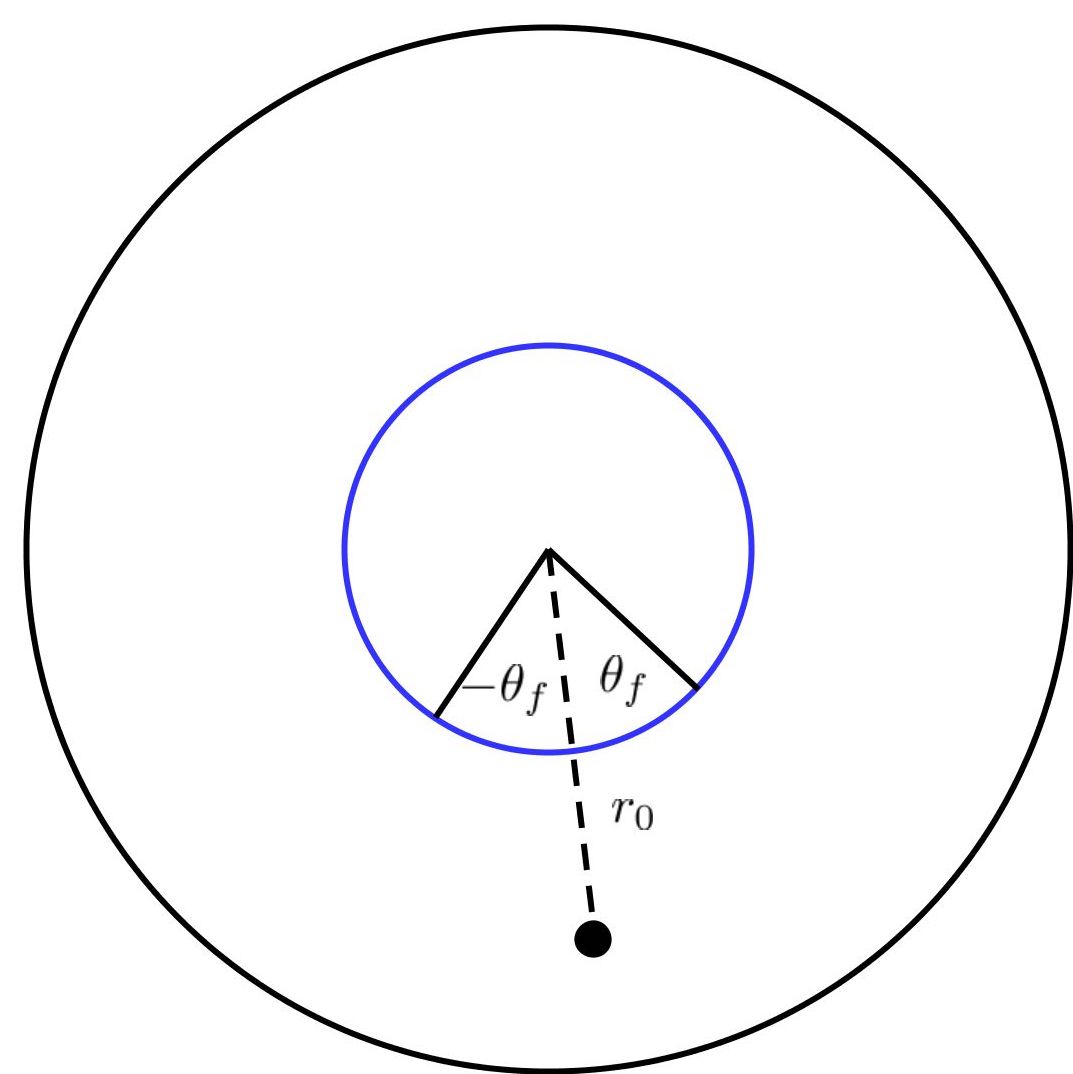} \\
\begin{center}
 \hspace{1.9cm}   (a) \hspace{5cm}     (b) \hspace{3.5cm}     (c)
\end{center}
\caption{General Channel Model for a point transmitter situated at $r=r_0$, a cylindrical absorbing receiver with radius $d_0$ and height $h$ surrounded by a larger cylindrical reflecting walls with radius $D_0$ and height $h$ (a). Note that the diffusion of the molecules is confined inside a finite annular volume as the z-dependence is suppressed through symmetry arguments (b). The receiver is modified to count only the particles inside the angle range $(-\theta_f,\theta_f)$  (c).}
\label{fig1}
\end{figure*}

\section{System Model}
\label{simMod}

The system model is depicted in Fig.\ref{fig1}. As indicated in the figure, a coaxial cylindrical absorbing receiver is placed at the center of the microfluidic channel whose boundaries are reflecting and a point transmitter transmits messages by releasing molecules to this receiver. Assuming that there is no flow in the environment, the movement of the released molecules are modelled by Brownian Motion as
\begin{subequations} \label{brown}
\begin{align}
  &\triangle x\sim \mathcal{N}\left( 0,2D\triangle t \right),\\
  &\triangle y\sim\mathcal{N}\left( 0,2D\triangle t \right),\\
  &\triangle z\sim\mathcal{N}\left( 0,2D\triangle t \right),
\end{align}
\end{subequations}
where $D$ is diffusion coefficient, $\triangle x$, $\triangle y$ and $\triangle z$ are the incremental step sizes in the three dimensions, $\triangle t$ is time step, and $ \mathcal{N}(\mu,\sigma^2)$ is the normal distribution with mean $\mu$ and variance $\sigma^2$. The cylindrical receiver absorbs the molecules that come to the  vicinity of its receptors and makes a decision by counting these absorbed molecules.
If the heights of the channel and the receiver $h$ are the same, then this system can be reduced to a 2-D bounded environment that has a concentric absorbing receiver and reflecting boundaries.  In practice, as in the case of oval cells, the length of the receiver can be smaller than the length of the channel. Therefore, we shall evaluate the required condition for the height of the receiver $h$ and height of the position of the transmitter $h_0$ for reducing this system to 2-D. The distribution of the released molecules along the $z$-axis can be modelled using (\ref{brown}) as $\mathcal{N}\left( h_0,2Dt \right)$. Therefore, if the arrival probability of any molecule at the bases of the receiver is almost 0, then our reduction is still valid. In other words, the microfluidic channel with coaxial cylindrical receiver can be reduced to 2-D if the following condition
\begin{equation}
     P\left( z<0 \right) +P\left( z>h \right) <\epsilon
\end{equation}
is satisfied. Then, using the distribution of $z$, we can write this condition explicitly as
\begin{equation}
 Q\left( \cfrac { h-{ h }_{ 0 } }{ \sqrt { 2Dt }  }  \right) +Q\left( \cfrac { { h }_{ 0 } }{ \sqrt { 2Dt }  }  \right) <\epsilon,
\end{equation}
where $t$ represents the maximum time of interest. Taking $h_0=h/2$, one can arrive at
\begin{equation}
 2Q\left( \cfrac { h }{2 \sqrt { 2Dt }  }  \right)  <\epsilon.
\end{equation}
Therefore, it can be concluded that if $h>>2\sqrt { 2Dt }$, then the system can be reduced to 2-D. We shall only discuss such systems and leave a more general analysis as a future work which shall be carried out using simulations rather than analytical derivations.

\section{Channel Impulse Response}
\label{secc}
Before performing the mathematical derivations, we first invoke a symmetry argument. As the receiver absorbs molecules at every height $z$ and any molecule for which $z<0$ or $z>h$ is reflected through the boundaries, the z-dependence of the channel can be suppressed. This is equivalent to a system model consisting of a 2-D annular channel and a point transmitter, as depicted in Fig. \ref{fig1}(b). Therefore, in order to derive the impulse response of the channel, we need to derive the probability density function of the molecules in the annular 2-D channel. To describe the diffusion of the molecule inside the annular region, we shall find a solution to the Fick's Law, satisfying the necessary boundary conditions 
\begin{equation} \label{eq:fick}
D \triangle \Prob = \frac{\partial \Prob}{\partial t},
\end{equation}
where $\triangle$ is the Laplacian operator and $\Prob$ is the probability density function (PDF) of the molecules inside the diffusion channel. The circular boundaries at $r=D_0$ are reflecting. Furthermore, the transmitter is assumed to be situated at a distance $r=r_0$ from the origin. In this section, we are interested in the absorption probability (an angle-independent quantity) of the molecules by the receiver. Therefore, our calculations include an SO(2) (angular) symmetry. In Section IV, the angle of absorption will be of interest; hence, we shall remove the SO(2) symmetry assumption. Finally, the probability distribution $\Prob$ should be zero when the molecules hit the receiver (assuming a perfect receiver due to simplicity), which results in the boundary conditions given as
\begin{subequations}  \label{eq:boundary}
\begin{align}
\frac{\partial \Prob}{\partial r}\Big|_{r=D_0} &= 0, \label{eq:Neuman} \\ 
 \Prob\Big|_{r=d_0} &= 0\label{eq:dirichlet},\\
P(r,0|r_0)&=\frac{1}{2\pi r} \delta(r-r_0),\label{eq:dirac}
\end{align}
\end{subequations}
where we recall that, since the boundaries are described by both Neumann and Dirichlet boundary conditions, the Laplacian operator is guaranteed to have a unique solution. 

We shall start with the separation of variables Ansatz defined as
\begin{equation*}
\Prob=\phi(r,\theta) T(t),
\end{equation*}
which leads to the equation
\begin{equation*}
D \frac{\Laplace \phi(r,\theta)}{\phi(r,\theta)} = \frac{T'(t)}{T(t)}= - \mu^2,
\end{equation*}
from which we can easily deduce 
\begin{equation*}
T(t) = e^{-\mu^2 t}
\end{equation*}
and arrive at the eigenvalue problem for Laplacian operator as
\begin{equation*}
 \Laplace \phi(r,\theta) = - \frac{\mu^2}{D} \phi(r,\theta).
\end{equation*}

As the boundary conditions are given by either Neumann or Dirichlet conditions, the eigenvalues $ \mu^2/D$ are non-negative and real, as well as eigenvectors corresponding to the distinct eigenvalues are orthogonal and form a basis for all possible solutions \cite{geometrical}. Here, we invoke the idea of SO(2) symmetry in our system. Due to angular symmetry, the position-dependent part of the Ansatz depends only on the distance from the origin and not the angle, i.e.  $\phi(r,\theta)=\phi(r)$. This choice eliminates certain eigenvalues (and corresponding eigenvectors) from the solution. Nonetheless, the coefficients corresponding to non-symmetrical eigenvectors are zero due to the symmetry of the system, removing our burden for further calculations.

Rewriting the eigenvalue equation in polar coordinates, we obtain
\begin{equation*}
r^2\phi''(r)+ r\phi'(r) + \frac{\mu^2}{D} r^2 \phi(r)=0,
\end{equation*}
which has the most general solution
\begin{equation*}
\phi(r) = c_1 J_0\left(\frac{\mu}{\sqrt{D}} r\right) + c_2 Y_0\left(\frac{\mu}{\sqrt{D}} r\right)\hspace{-0.3em},
\end{equation*}
where $J_n$ and $Y_n$ are the Bessel functions of the first and second kind, respectively. We are now ready to shape our solution according to the boundary conditions given in (\ref{eq:boundary}).

At this point, we shall stress that we cannot dispose off $Y_n$, since we do not require a solution for $r=0$, i.e. $r=0$ is not in the domain of the analytical function that we are interested in.  We shall define the special function $\eta_0(\beta_n x)$ as
\begin{equation*}
\eta_0(\beta_n x) =  J_0\left(\beta_n x\right) + c_n Y_0\left(\beta_n x\right)
\end{equation*}
such that $\eta_0'(\beta_n)=- \eta_1(\beta_n) =0$ and $\eta_0(\alpha \beta_n)=0$, where $\alpha=d_0/D_0$ and $\eta_0'(x)$ denote the derivative of $\eta_0(x)$ with respect to $x$. Similarly, $\eta_m(\beta_n x)$ is defined with the corresponding Bessel functions $J_m$ and $Y_m$ and the same coefficients $\beta_n$ and $c_n$. We shall discuss the construction of such a function in Section V. For now, we note that $\{\beta_n\}$, called as eigenvalues from now on, is an (increasingly) ordered, discrete, and infinite set. It is verified through straightforward algebra that the function $\phi(r)=\eta_0(\beta_n r/D_0)$ satisfies the two boundary conditions and is a radial solution for the diffusion equation given in (\ref{eq:fick}). The following orthogonality condition can be shown to hold for $\eta_0(\beta_n x)$:
\begin{align*}
\int_{\alpha}^1 \eta_0(\beta_n x) \eta_0(\beta_m x) x \diff x = \frac{1}{2}\left( \eta_0^2(\beta_n) - \alpha^2 \eta_1^2(\alpha \beta_n)   \right) \delta_{nm}.
\end{align*}

Bringing the radial $\phi(r)$ and time $T(t)$ solutions together, we find the most general to be of the form
\begin{equation*}
\Prob = \sum_{n=1}^\infty A_n \eta_0\left(\beta_n \frac{r}{D_0}\right) e^{- \beta_n^2 \frac{D t}{D_0^2}},
\end{equation*}
where we note that $\beta_1 >0$ (see Table 1 in Appendix), hinting that the final probability density will be zero everywhere in space. Taking the orthogonality condition into account, we can find the general normalization constant $A_n$ as
\begin{align*}
A_n &= \frac{1}{\pi D_0^2} \eta_0\left(\beta_n \frac{r_0}{D_0}\right) \frac{1}{\left( \eta_0^2(\beta_n) - \alpha^2 \eta_1^2(\alpha \beta_n)   \right)},
\end{align*}
from which we find the solution to be
\begin{equation} \label{eq:final}
\begin{split}
\Prob =\sum_{n=1}^\infty & \frac{ \eta_0\left(\beta_n \frac{r_0}{D_0}\right)\eta_0\left(\beta_n \frac{r}{D_0}\right)}{\pi D_0^2\left( \eta_0^2(\beta_n) - \alpha^2 \eta_1^2(\alpha \beta_n)   \right)}    e^{- \beta_n^2 \frac{D t}{D_0^2}},
\end{split}
\end{equation}
where we recall that $\{\beta_n\}$ are defined such that $\eta_0(\alpha\beta_n )=0$ and $\eta_1(\beta_n)=0$ to satisfy boundary conditions. Now that we have the PDF $\Prob$, we can calculate the hitting number as
\begin{equation*}
n_{hit}(t) =2 \pi d_0 D \frac{\partial \Prob}{\partial_r}\Big|_{r=d_0},
\end{equation*}
 where $D \frac{\partial \Prob}{\partial_r}\Big|_{r=d_0}$ represents the probability current into the absorbing receiver. From the probability density function given in (\ref{eq:final}), we find the hitting number to be
 \begin{equation} \label{eq:nhit}
\begin{split}
n_{hit}(t) =- 2  D \sum_{n=1}^\infty & \frac{\alpha \beta_n  \eta_0\left(\beta_n \frac{r_0}{D_0}\right)}{ D_0^2\left( \eta_0^2(\beta_n) - \alpha^2 \eta_1^2(\alpha \beta_n)   \right)} \eta_1\left(\beta_n \alpha \right) e^{- \beta_n^2 \frac{D t}{D_0^2}}.
\end{split}
 \end{equation}
 
  \begin{figure*}
\centering
\includegraphics[width=8cm,height=4.5cm]{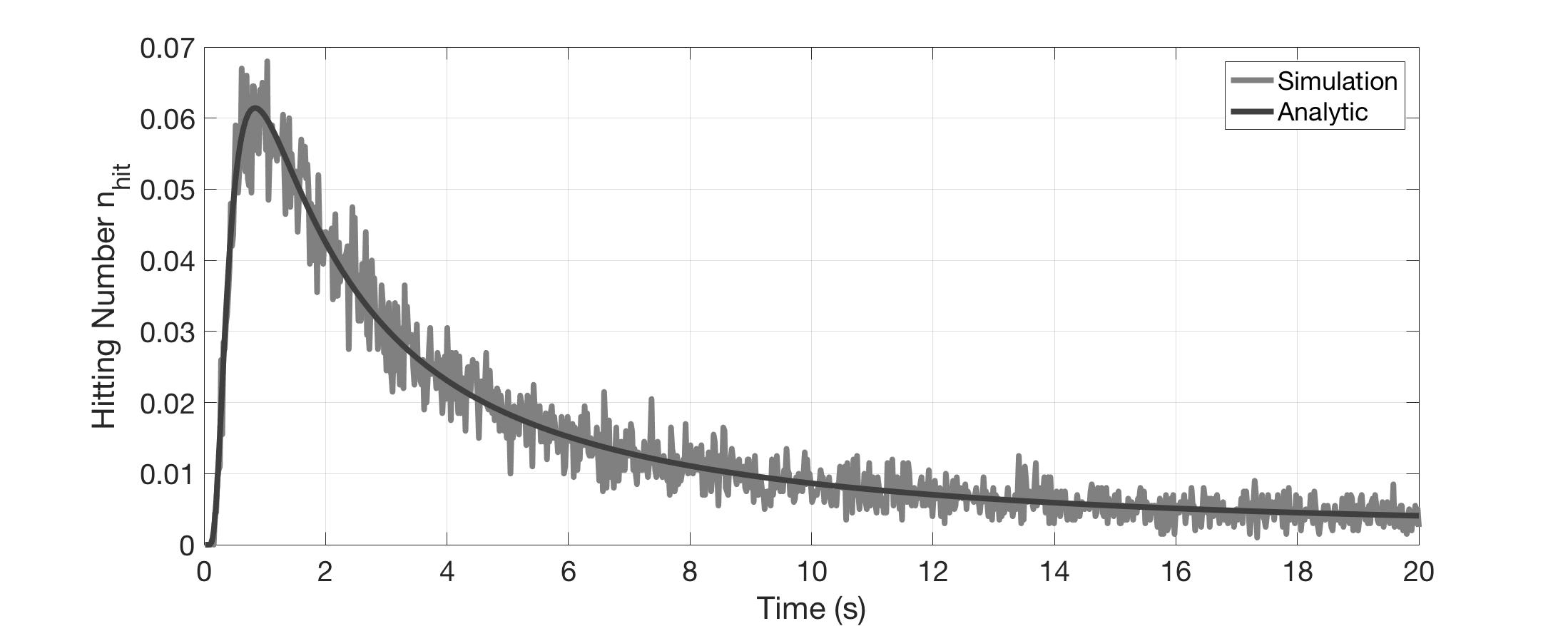} 
\includegraphics[width=8cm,height=4.5cm]{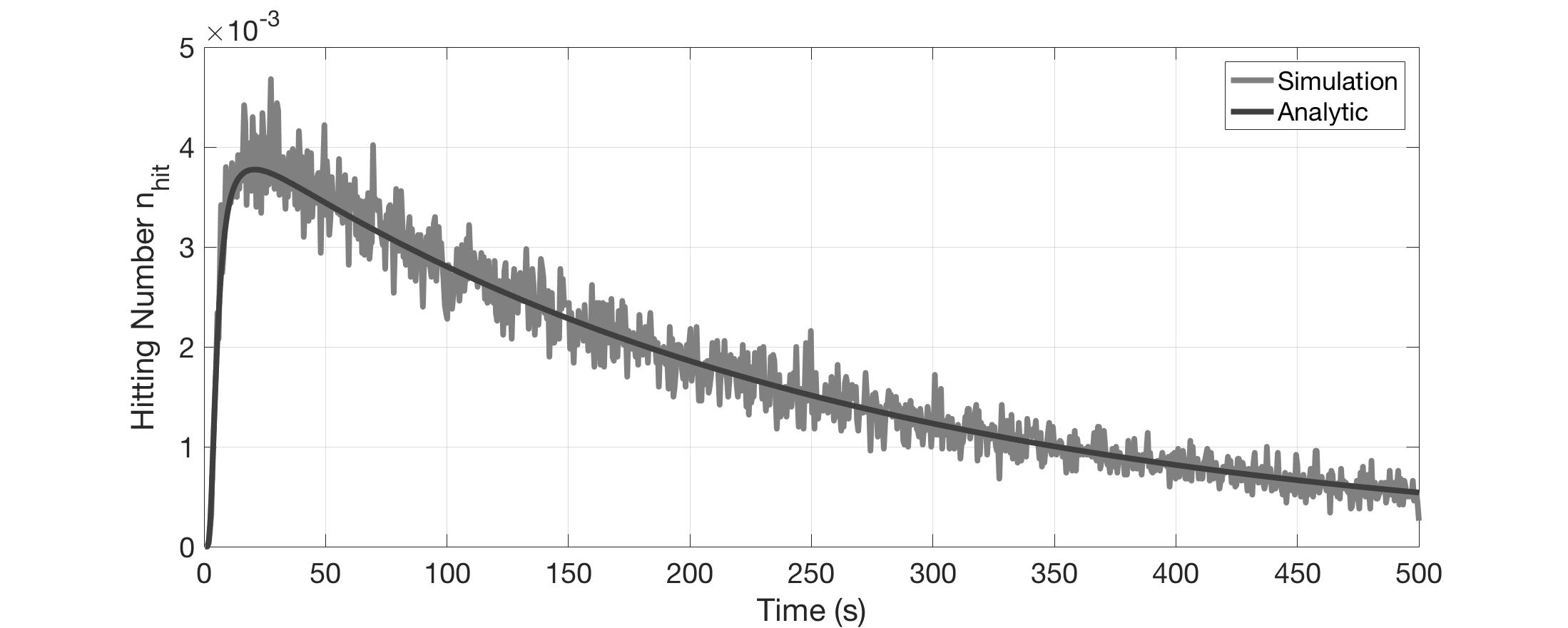} \\
\begin{center}
    (a) \hspace{9cm} (b)
\end{center}
\includegraphics[width=8cm,height=4.5cm]{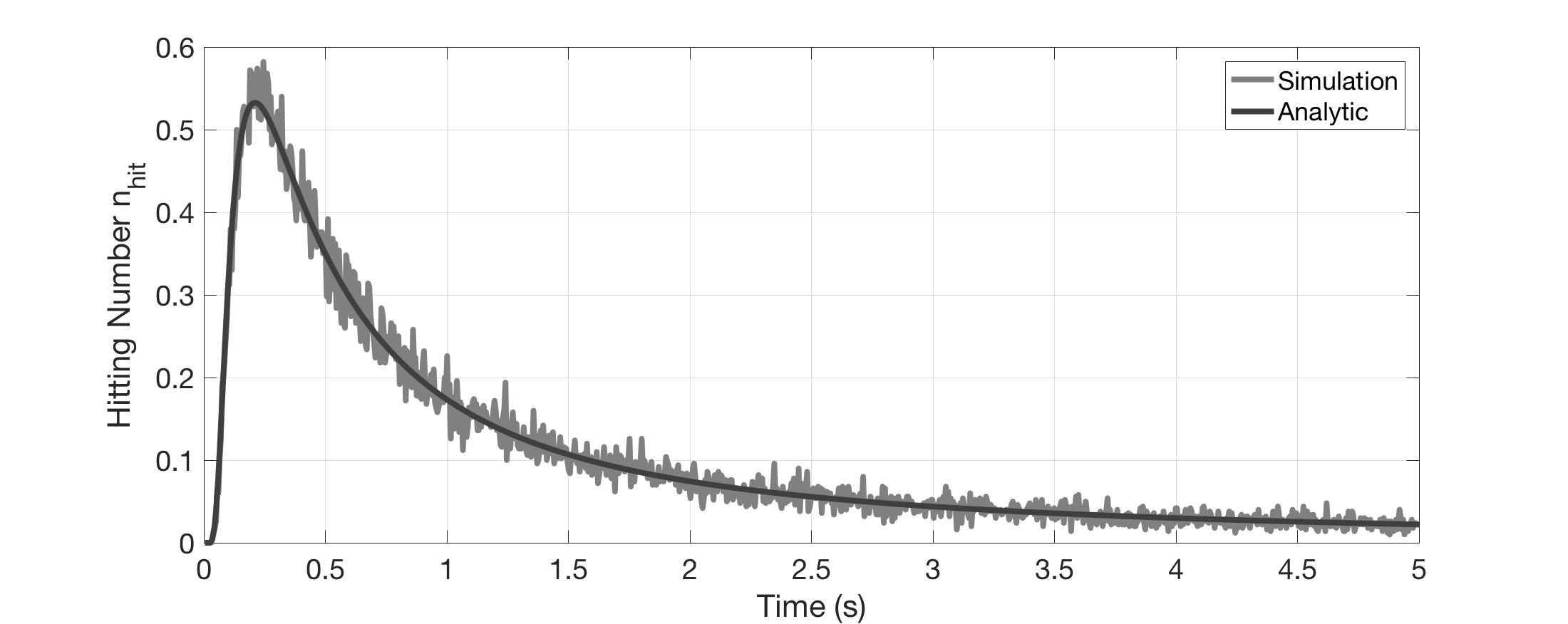} 
\includegraphics[width=8cm,height=4.5cm]{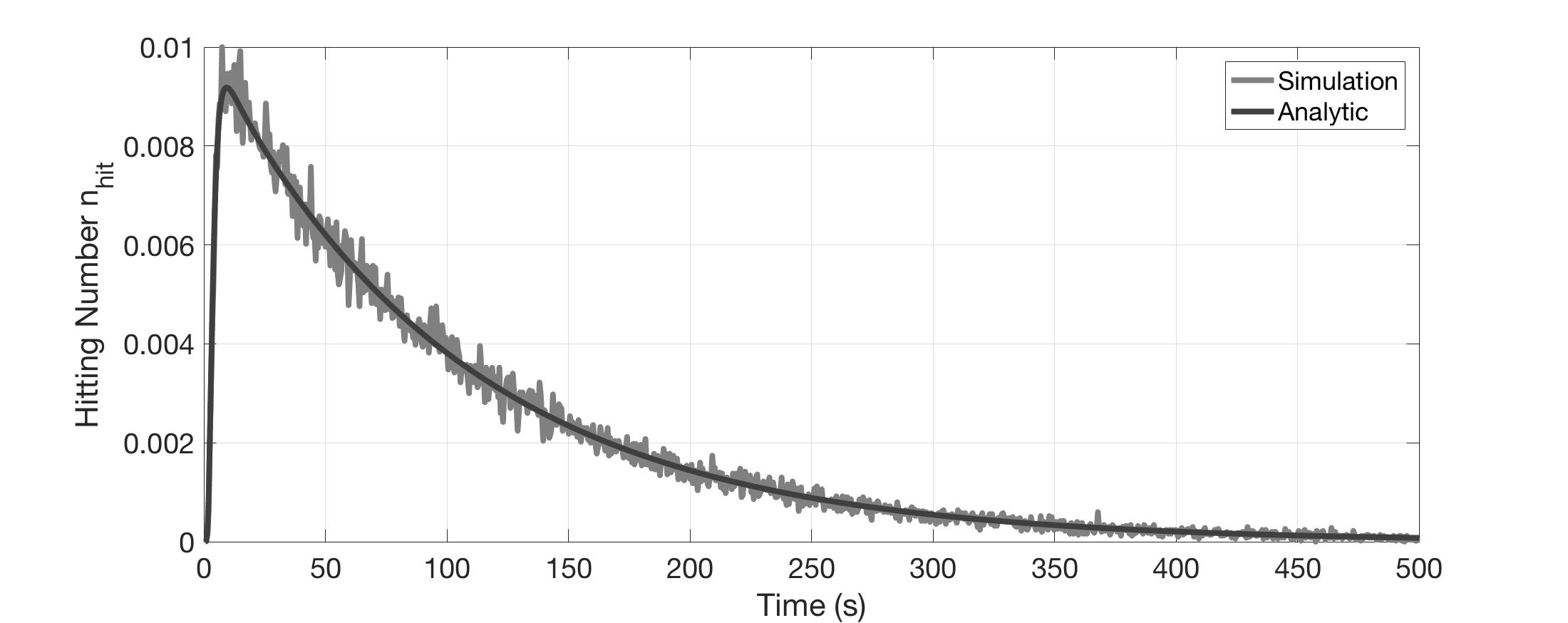}\\
\begin{center}
    (c) \hspace{9cm} (d)
\end{center}
\caption{Simulations of hitting number $n_{hit}(t)$ versus time for $D= 80 \mu m^2/s$, $D_0=100 \mu m$, $d_0 = 1 \mu m$ (a,b) and $d_0=10 \mu m$ (c,d), $r_0=20 \mu m$ (a,c) and $r_0=70 \mu m$ (b,d). Note the correspondence between the analytical solution and the simulation in each case. } \label{fig:hitting}
\end{figure*}
 \noindent
 Moreover, having found the PDF $\Prob$, we can find the radial distribution $\Probs$ as
 \begin{equation*}
     \Probs = 2\pi r \Prob.
 \end{equation*}
 
 In this article, we mainly discuss the hitting number $n_{hit}(t)$; nonetheless, the radial distribution can be useful for modelling the behavior of the molecules in the presence of a flow for further analysis. 

\section{Angular Dependent Channel Impulse Response}

Inspired from the nature of diffusion, it is shown that using a partially-counting receiver based on angular position has beneficial effects in molecular communications \cite{akdeniz2017molecular}. Since molecules move slowly, it takes much higher expected time to move to the part of the receiver, which are far from the transmitter. These parts can also be represented by the reception angle as shown in Fig. \ref{fig1}(c) and angle-dependent channel impulse response can be used to improve the channel performance by reducing inter symbol interference as proposed in \cite{akdeniz2017molecular}.

The receiver (in Fig. \ref{fig1}(c)) absorbs all the molecules incident upon itself, but counts only those that arrive inside the angle interval $[-\theta_f,\theta_f]$ and disregards the rest. We can modify our previous calculations to find an analytical solution for this case as well, which is carried out in the Appendix. 

For this channel, we define the hitting number as probability of a  single released molecule to hit the receiver inside the angle range $[-\theta_f,\theta_f]$ between times $t$ and $t+\diff t$
\begin{align*}
    n_{hit}(\theta_f,t) = \int_{-\theta_f}^{\theta_f} D d_0 \partial_r \Prob |_{r=d_0}.
\end{align*}

From (\ref{eq:probfinal}), we find the hitting number as
\scriptsize
\begin{align} \label{eq:hitangle}
    &n_{hit}(\theta_f,t)= \sum_{ n=1}^\infty \frac{\theta_f D \alpha \beta_{0n}}{ \pi D_0^2 I_{0n}} \eta_0\left(\beta_{0n} \frac{r_0}{D_0} \right) \eta_0 \left(\beta_{0n} \frac{d_0}{D_0}\right)' e^{-\beta_{0n}^2 \frac{Dt}{D_0^2}}+ \nonumber \\
&\sum_{m=1, n=1}^\infty \frac{2D\alpha \beta_{mn} }{m\pi D_0^2 I_{mn}} \sin(m \theta_f) \eta_m\left(\beta_{mn} \frac{r_0}{D_0} \right) \eta_m \left(\beta_{mn} \frac{d_0}{D_0}\right)' e^{-\beta_{mn}^2 \frac{Dt}{D_0^2}}
\end{align}
\normalsize

\begin{figure*}
     \includegraphics[width=9cm]{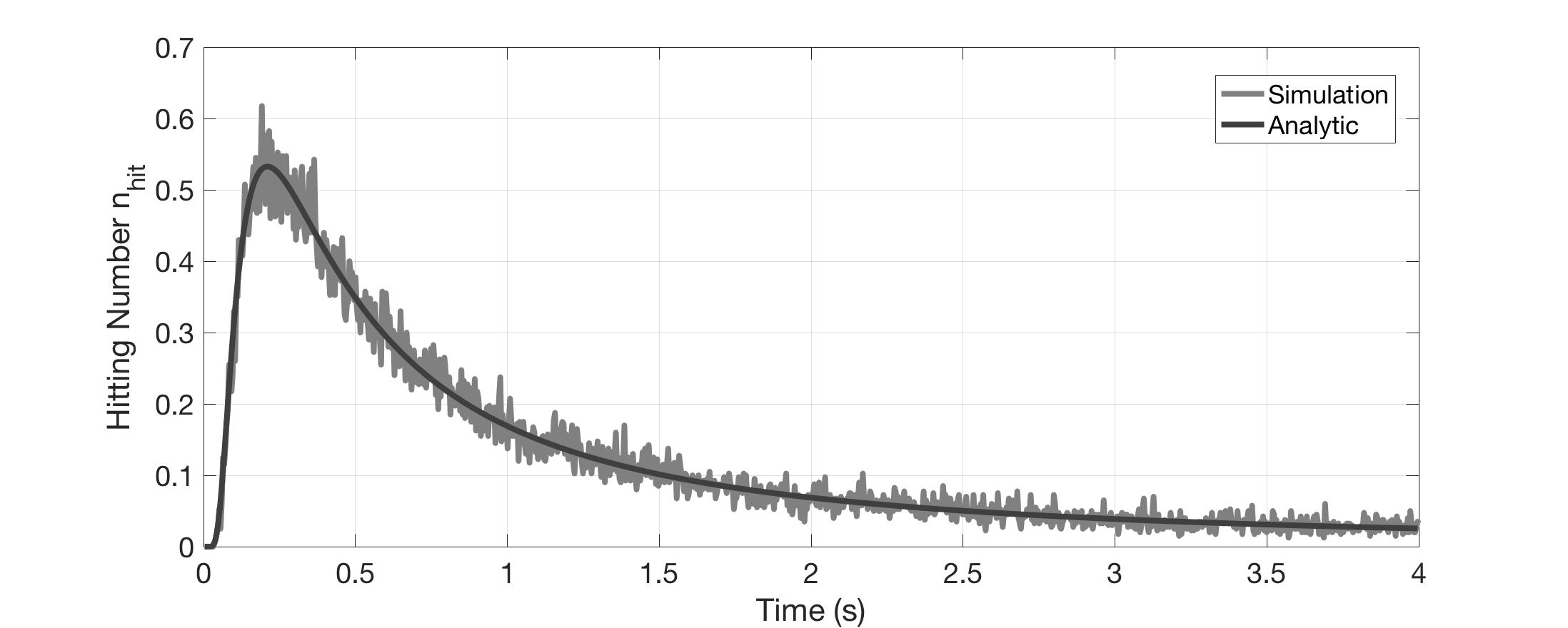}
     \includegraphics[width=9cm]{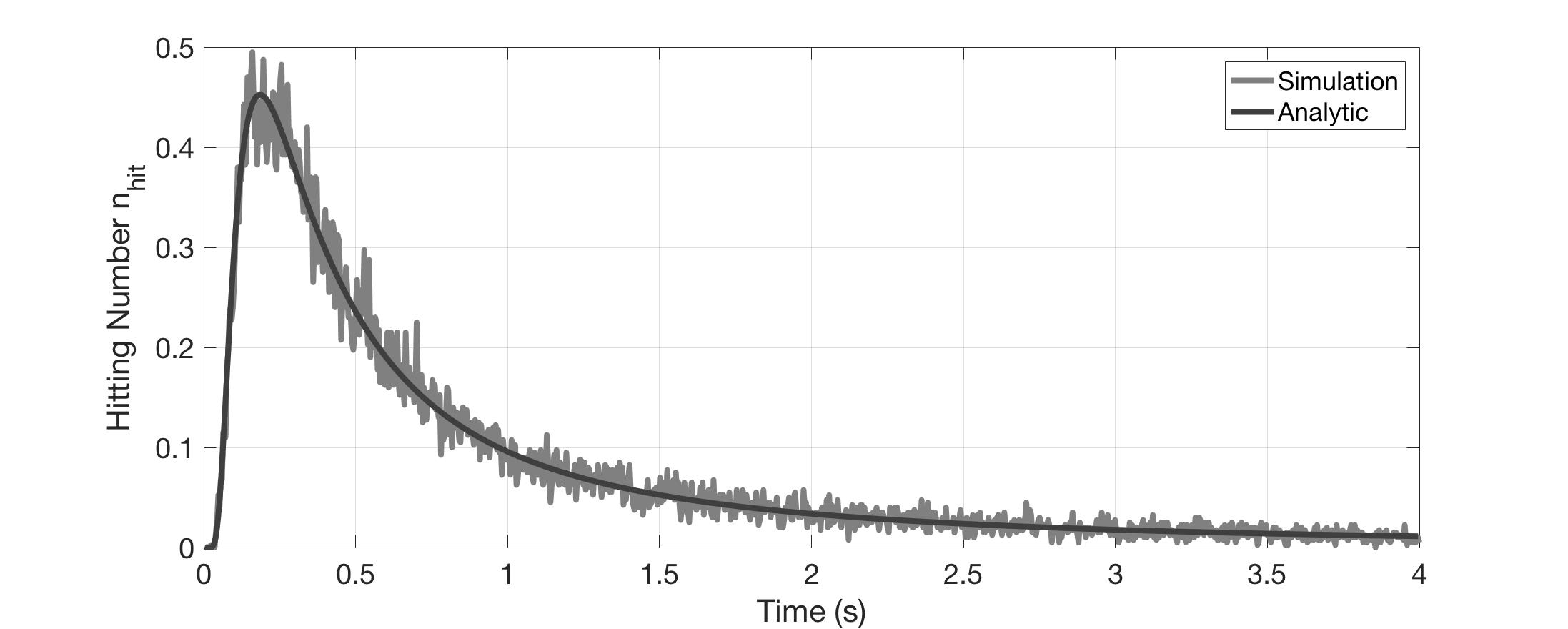}
\begin{center}
    (a) \hspace{9cm} (b)
\end{center}
     \caption{Comparison of Simulation and the Angle Dependent Analytical Solution for $D_0 = 100\mu m$, $d_0=10\mu m$, $r_0=20 \mu m$, $D=80 \mu m^2/s$, $\theta_f = \pi /2$ (a) and $\theta_f = \pi/6$ (b). Comparing with Fig. \ref{fig:hitting}(c), one can see that for $r_0=20 \mu m$ almost all particles hit inside the angle range $[- \pi/2,\pi/2]$.   } \label{fig:theta}
 \end{figure*}

Analytical and simulation result comparison for this channel type is given in Fig. \ref{fig:theta}. For the scope of this paper, we focus on the receiver type with $\theta_f = \pi$, for which (\ref{eq:hitangle}) reduces to (\ref{eq:nhit}).

\section{Numerical Example and Comparison with the Simulation}
Having found the analytical solution for the 2-D annular channel, we shall now focus on verifying our findings through comparison with a simulation, and then discuss the effects of reflecting boundary on the channel response. In this section, we first describe the simulation model, derive the numerical solutions for $\eta_0(\beta_n x)$, then simulate the hitting number $n_{hit}(t)$ for different aspect ratios $\alpha=\frac{d_0}{D_0}$, and finally define and interpret certain channel characteristics.



In our simulations, we take the radius of the outer cylinder as $D_0=100 \mu m$ and simulate the system for the different receiver radii $d_0$ by changing $\alpha=\frac{d_0}{D_0}$. Our channel is controlled by a diffusion process, and we model it with a diffusion coefficient $D=80$ $\mu m^{2}/s$.

\subsection{Derivation of $\eta_0(\beta_n x)$ }
When finding the impulse response of the 2-D channel, we have assumed that there exist functions $\eta_0(\beta_n x)$ such that $\eta_0'(\beta_n)=0 $ and $\eta_0(\alpha \beta_n )=0$. In this section, we shall discuss how to construct such functions and illustrate an algorithm to find $\beta_n$'s.

To begin with, we can rearrange the radial solution slightly differently, ignoring the general normalization constant for now as
\begin{equation*}
\phi(r) =  J_0\left(a r\right) + c Y_0\left(a r\right),
\end{equation*}
where we define $a=\frac{\mu}{\sqrt{D}}$ for simpler algebra. Using the boundary conditions (\ref{eq:Neuman}) and (\ref{eq:dirichlet}), we find the following set of linear equations:
\begin{subequations} 
\begin{align*}
- a J_1(aD_0) - c a Y_1(aD_0) &= 0, \\
J_0(ad_0) + cY_0(ad_0)&=0.
\end{align*}
\end{subequations}
Rearranging the terms, we can obtain
\begin{subequations} 
\begin{align*}
c  &= - \frac{J_1(aD_0)}{Y_1(aD_0)}, \\
c  &= - \frac{J_0(ad_0)}{Y_0(ad_0)},
\end{align*}
\end{subequations}
where setting $aD_0=\beta$ and $\alpha=\frac{d_0}{D_0}$, we  look for the solutions of the equation
\begin{align*} \label{eq:num}
\frac{J_1(\beta)}{Y_1(\beta)}-\frac{J_0(\alpha \beta)}{Y_0(\alpha \beta)}=0,
\end{align*}
which we call the \textit{characteristic equation}.
There are infinitely many solutions for this equation, each of which corresponds to a distinct eigenvalue and an eigenfunction of the Laplacian operator. We also note that $c$ is fully determined by the procedure above. Finally, we finish our derivations by defining the function
\begin{equation*}
\eta_0(\beta_n x) =  J_0\left(\beta_n x\right) + c_n Y_0\left(\beta_n x\right).
\end{equation*}

Without a given aspect ratio $\alpha$, this is the most general function we can define. If $\alpha$ is given, then we can construct a code that finds the roots of the characteristic equation. This is feasible, because the roots are inside certain periodic intervals even though they are not periodic. Once the roots $\beta_n$ are found, we can construct the eigenvectors $\eta_0(\beta_n x)$ by finding $c_n$'s.

\subsection{Hitting Number Comparison}

Now that we have constructed both our analytical solution and the simulation results, we shall compare the hitting number, $n_{hit}(t)$, for different aspect ratios, $\alpha$, and different initial positions, $r_0$, in Figure \ref{fig:hitting}. As can be seen from the figure, the simulation and the analytical function are in agreement for multiple scenarios, as expected. 

To understand the channel performance, we propose the following definitions for different channel characteristics:
\begin{definition}[Peak time] 
The peak time $\tau_{peak}$ is defined as the time such that the hitting number is maximum, e.g.
\begin{equation*}
    \frac{\partial n_{hit}(t)}{\partial t}\Big|_{t=\tau_{peak}}=0.
\end{equation*}
\end{definition}
\begin{definition}[Average time]
The average time $\tau_{average}$ is defined as the expectation value of the time where the hitting number $n_{hit}(t)$ is taken to be the probability density function, i.e.,
\begin{equation*}
    \tau_{average}=<t>= \int_0^\infty  \, t  \, n_{hit}(t) \, \diff t.
\end{equation*}
Note that the hitting number being the probability density function for time is a direct consequence of the continuity equation.
\end{definition}
\begin{definition}[Half time]
The half time $\tau_{half}$ is defined as the time it takes for the molecule to be absorbed with a probability of $0.5$, i.e.,
\begin{align*}
\int_0^{\tau_{half}} n_{hit}(t) \diff t = 0.5.
\end{align*}
\end{definition}
 
One should keep in mind that we are required to find different eigenvalues for each aspect ratio $\alpha$ to construct the special functions $\eta_0(\beta_n x)$. Once the analytical solution is constructed, it can be shown to agree with the simulations as presented in this section.
 
 \begin{figure*}
     \includegraphics[width=9cm]{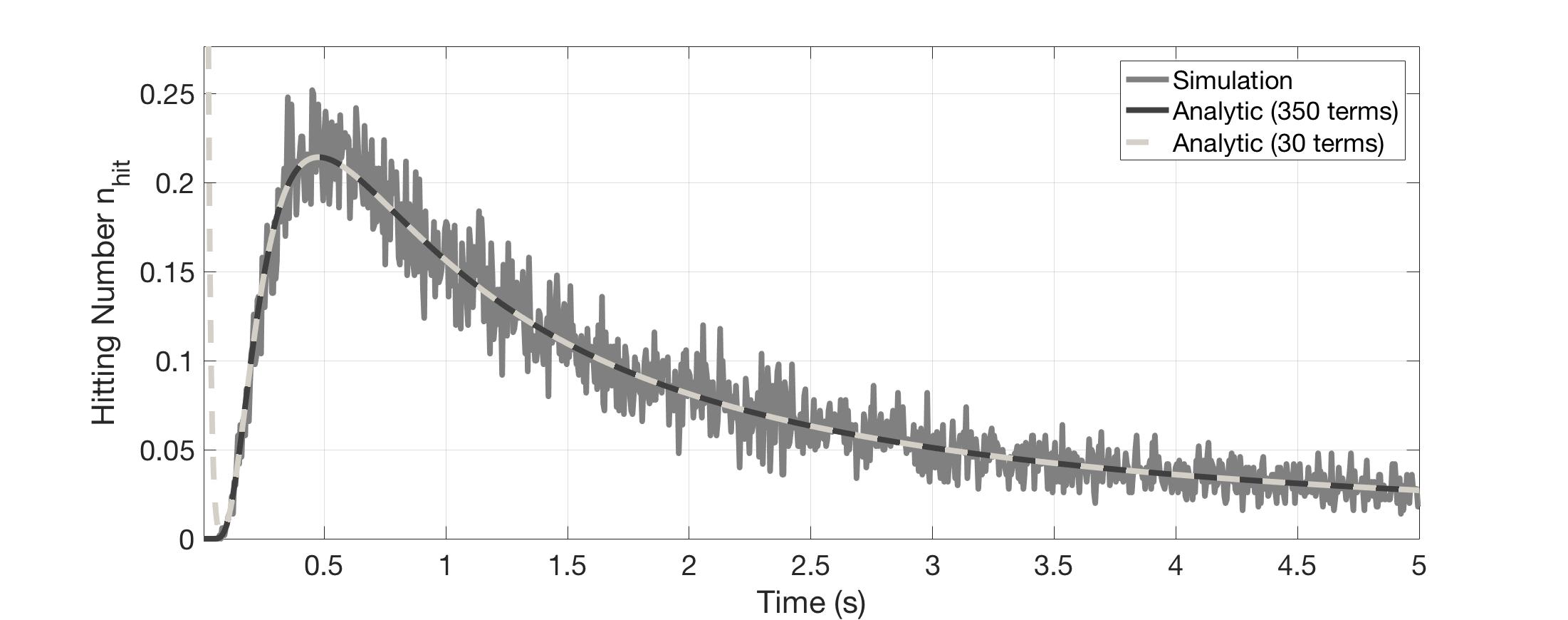}
     \includegraphics[width=9cm]{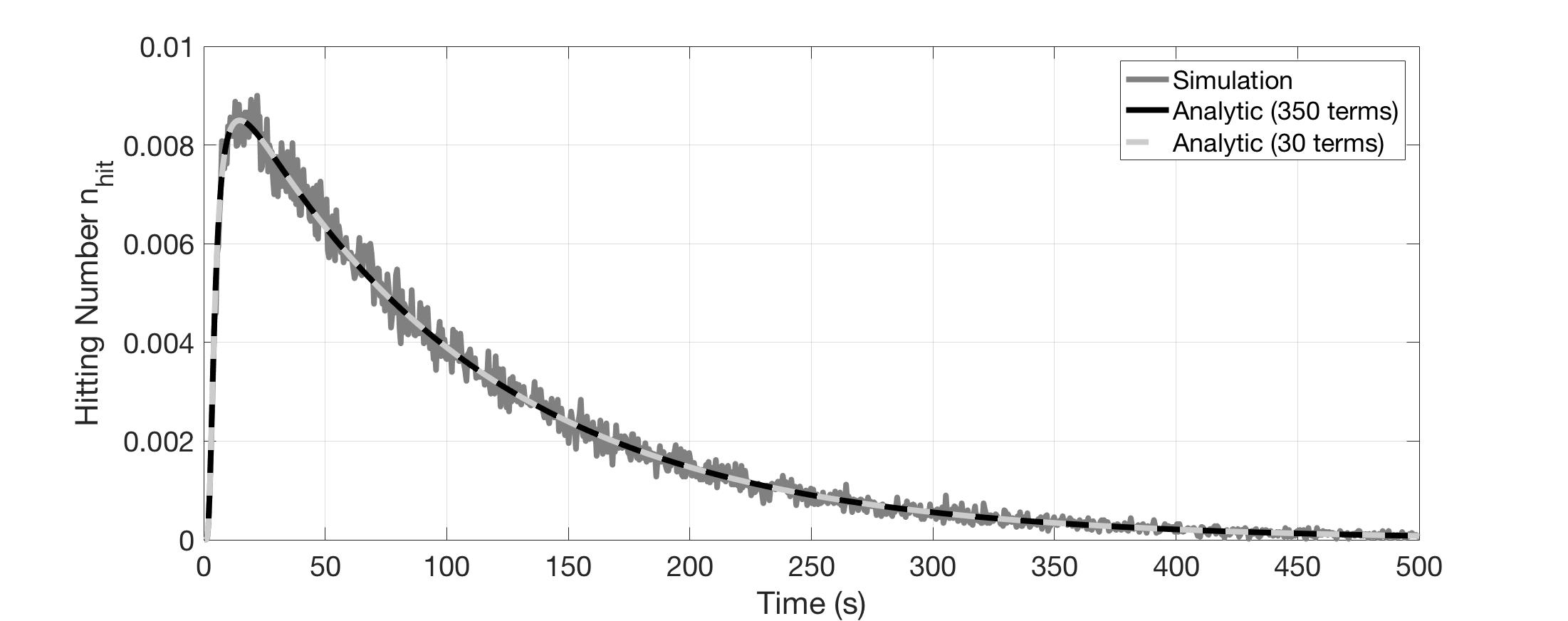}
\begin{center}
    (a) \hspace{9cm} (b)
\end{center}
     \caption{Comparison of Simulation and Analytical Solution for finite number of terms in (\ref{eq:nhit}) for $D_0 = 100\mu m$, $d_0=10\mu m$, $r_0=25 \mu m$ (a) and $r_0 = 75 \mu m$ (b).} \label{fig:error}
 \end{figure*}
 
  \begin{figure*}
\centering
\includegraphics[width=8cm,height=4.5cm]{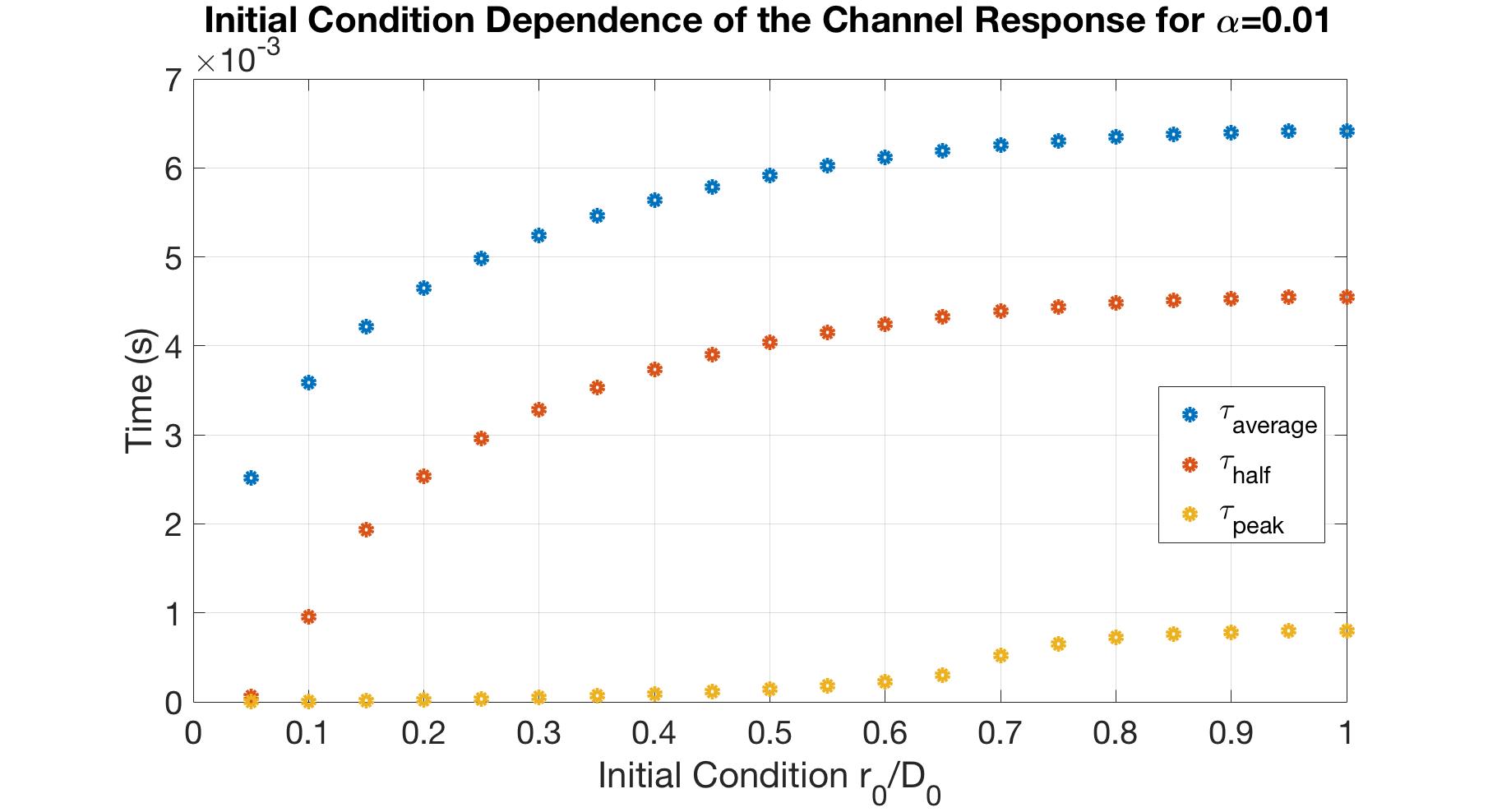} 
\includegraphics[width=8cm,height=4.7cm]{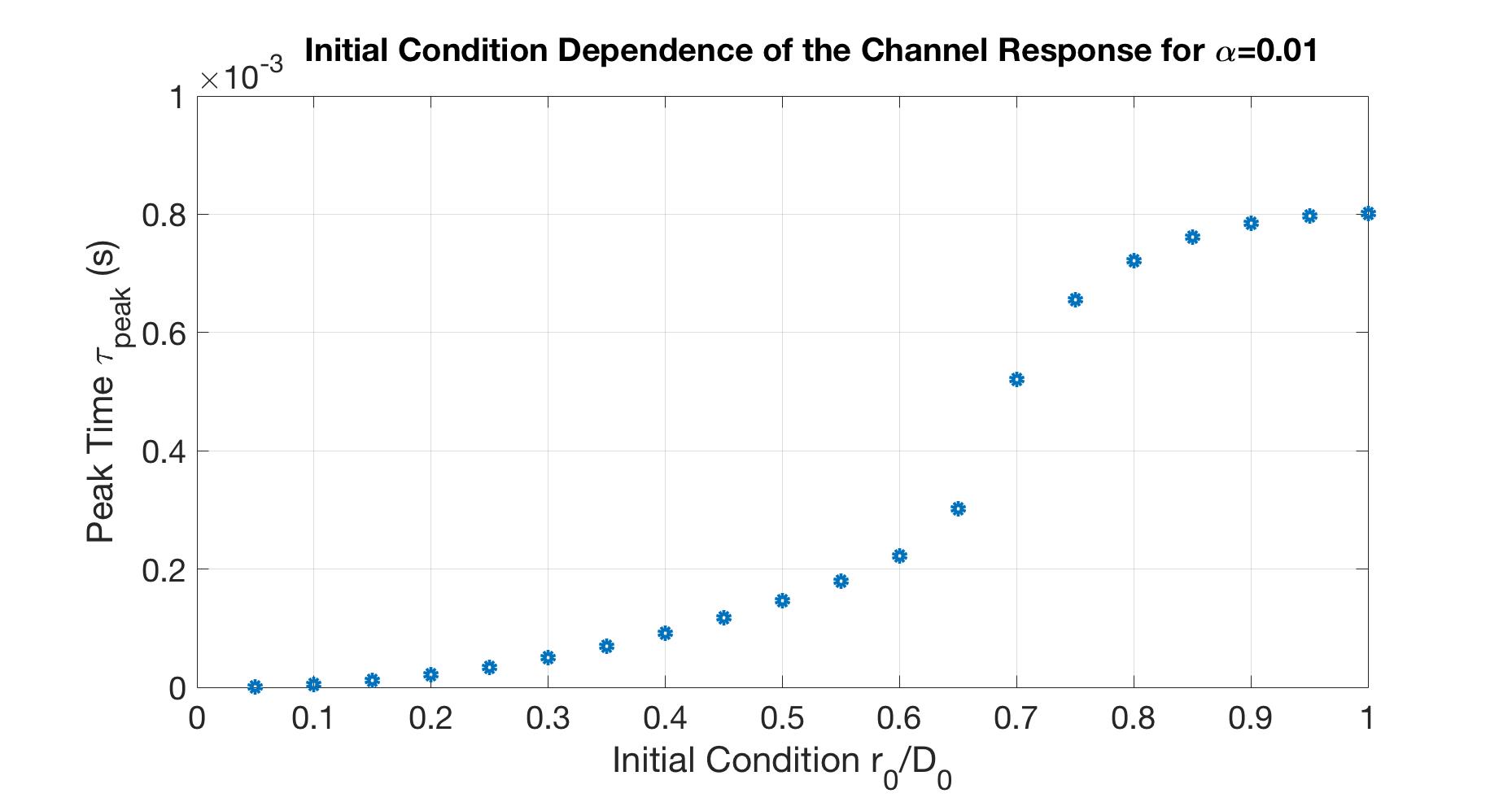} 
\begin{center}
    (a) \hspace{9cm} (b)
\end{center}
\includegraphics[width=8cm,height=4.5cm]{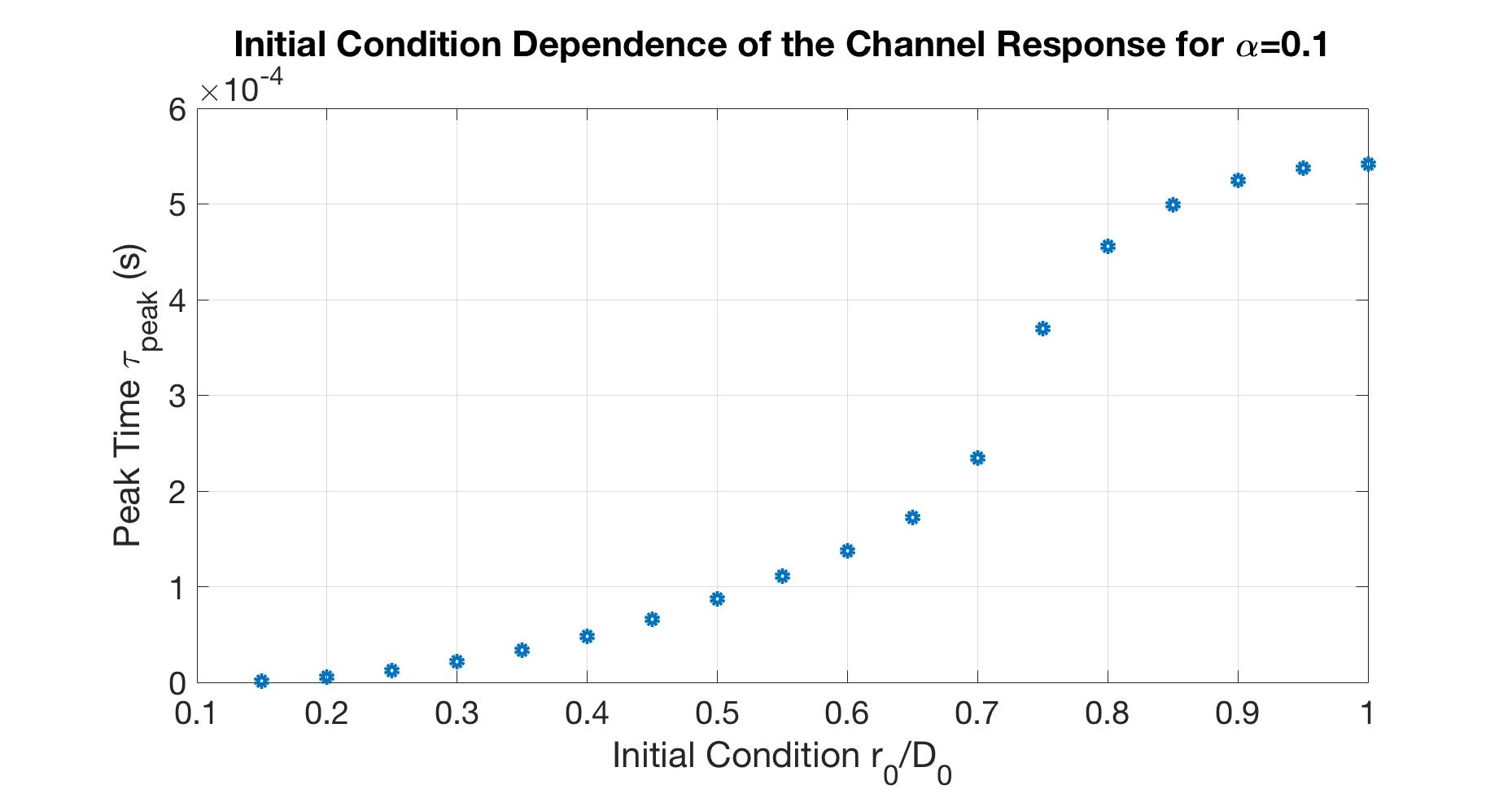} 
\includegraphics[width=8cm,height=4.5cm]{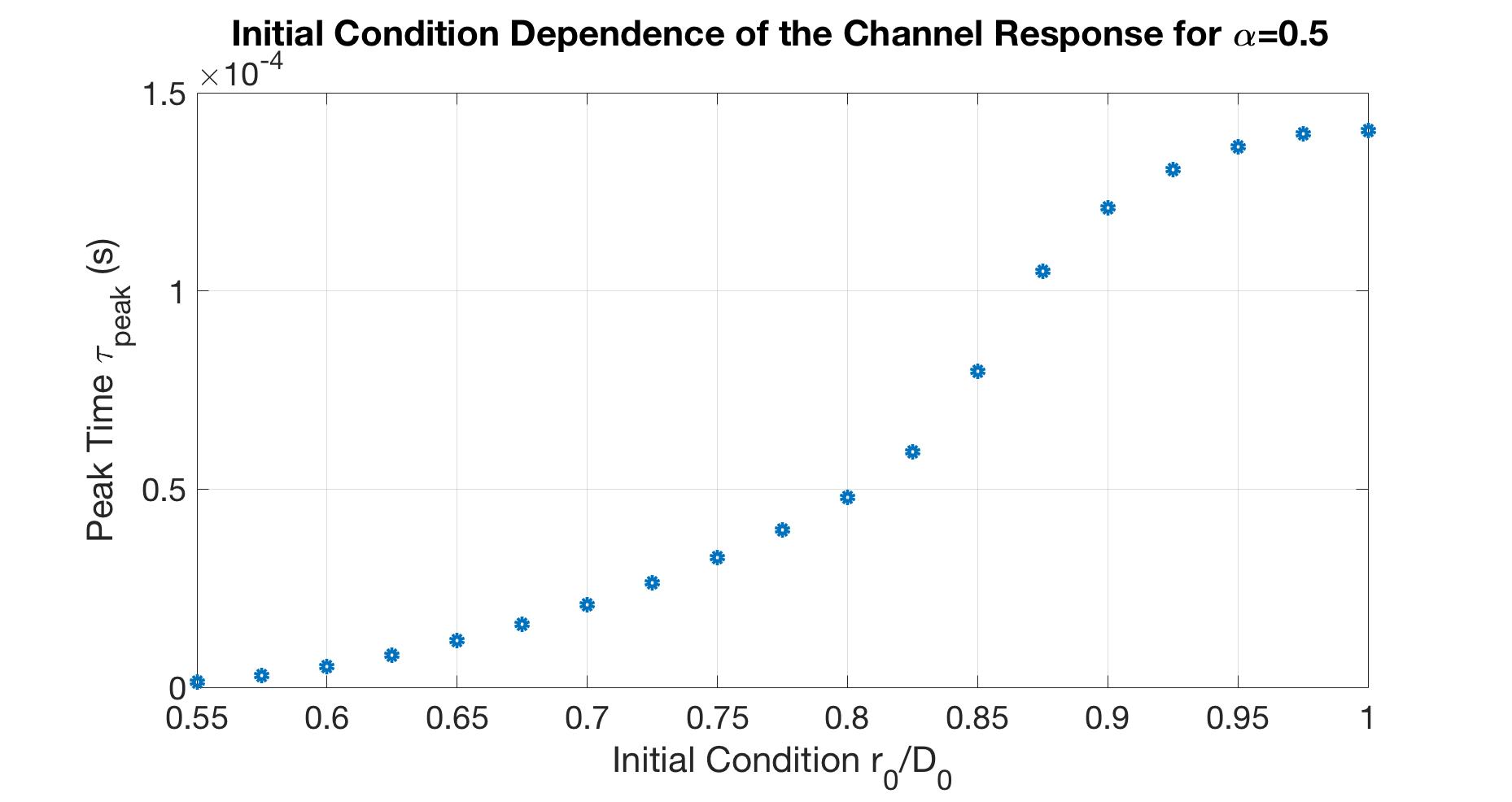} 
\begin{center}
    (c) \hspace{9cm} (d)
\end{center}
\caption{The channel characteristic times ($\tau_{average}$, $\tau_{half}$ and $\tau_{peak}$) for different aspect ratios $\alpha$. Note the apparent trend change for $\tau_{peak}$ once the initial release point $(r_0-d_0) \simeq \frac{2}{3} l_c$, where $l_c=D_0-d_0$ is the channel length. This is due to molecules reflecting from the boundary being dominant for the absorption. For closer initial release points $r_0$, the peak time vs. initial distance scales as $\tau_{peak} \sim (r_0-d_0)^2$ in agreement with the 3-D spherical receiver point transmitter case \cite{yilmaz2014three}. The channel characteristics are calculated and represented for $D_0=500 nm$ and $D=80 \mu m^2/s$. Nonetheless due to inherent space scaling symmetry, the shape of the curves are the same for micro-scales as well, with the exception of larger time values.}
\label{fig:meantime}
\end{figure*}

Moreover, the infinite sum presented in (\ref{eq:nhit}) can be practically terminated at a $\beta_N$-th term, as long as for the final term the condition 
\begin{equation}
    \exp(-\beta_N^2 \frac{D t}{D_0^2}) << 1,
\end{equation}
where $t$ is the time after which the simulation and the truncated analytical solutions are in agreement, is satisfied. Such comparison can be found in Fig. \ref{fig:error} for different number of terms of the summation.

\subsection{Peak-time, Average Time and Half-time Simulations}

The 2-D annular channel characteristics can be captured through the peak, average, and half-time values and their dependence on initial release point $r_0$, which can be seen in Fig. \ref{fig:meantime}. 

The effect of the reflecting boundary can best be seen from the peak time $\tau_{peak}$. When the molecule is initially close to the receiver, the effect of the boundary is negligible and we observe a square-law dependence between the distance and the peak-time $\tau_{peak}$, as was the case for a 3-D spherical receiver and a point transmitter. Defining the channel length $l_c = D_0 - d_0$, we realize that the deviation from the square-law is apparent when the release distance is $(r_0-d_0) \simeq \frac{2}{3} l_c$. The same transition is not as apparent with $\tau_{half}$ and $\tau_{average}$, as the existence of the reflecting boundary ensures that the molecule is eventually absorbed. Therefore, the effect of the boundary on these values is present even when the molecule is initially very far away from the boundary and close to the receiver. 

As there are infinitely many summed terms in the expression for $n_{hit}(t)$, it is unfortunately not straightforward to obtain a formula for $\tau_{peak}$ and $\tau_{half}$. Nonetheless, one can find the average time analytically and show through a comparison test that its analytical expression is, indeed, convergent:
\begin{equation}
    \tau_{average} = -2  \sum_{n=1}^\infty  \frac{\alpha D_0^2 \eta_0\left(\beta_n \frac{r_0}{D_0}\right) \eta_1\left(\beta_n \alpha \right) }{D \beta_n^3 \left( \eta_0^2(\beta_n) - \alpha^2 \eta_1^2(\alpha \beta_n)   \right)}.
\end{equation}

\section{Conclusion}

In our work, we derive the impulse response of a diffusion channel with point transmitter and coaxial cylindrical absorbing receiver first for SO(2) symmetric initial conditions and then breaking the symmetry while offering a more rigorous and angle dependent description for the impulse response inside the channel. Due to symmetry of the system in $z$ coordinates, we are able to reduce the channel behavior onto 2-D and find the impulse response of a 2-D annular channel with a point transmitter. In this pursuit, we define a special function $\eta_0(\beta_{n} x)$ (or $\eta_m(\beta_{mn}x)$ in general), which is a combination of Bessel functions of the first and second kind, for the impulse response to both satisfy the necessary boundary conditions and to be an exact solution to the diffusion equation. This method of obtaining an impulse response leads to an infinite number of terms, sum of which converge for $t>0$.  

Through corresponding Monte-Carlo simulations, it is shown that the infinite sum in the analytical solution can be truncated after certain number of terms depending on the time interval one is interested in. If an analytical solution is desired for even later times, less terms in the analytical solution will be required to accurately depict the behavior of the channel. Furthermore, we show the equivalence between the Monte-Carlo simulations and the analytical solutions for different channel and receiver parameters, such as the aspect ratio $\alpha$, the boundary radius $D_0$ and the receiver radius $d_0$. This equivalence lays evidence for the accuracy of our findings. 

Afterwards, we explore the dependency of certain channel characteristics on the initial position of the transmitter. As is clear from the peak time $\tau_{peak}$ simulations, the effect of the boundary on the peak time is more apparent as the initial position of the transmitter is around $2/3 l_c$, where $l_c=D_0-d_0$ is the channel length. As $r_0>l_c$, there is an apparent trend shift in the behavior of the peak time caused by the boundary. Nonetheless, this trend shift is not as apparent for the average and half time, $\tau_{average}$ and $\tau_{half}$, respectively. The intuitive reason behind this phenomenon can be explained through the tail effect. The average and half time values are more dependent on the existence of the boundary as they rely not only on the peak of the hitting number, but also on the behavior of the tail that follows the peak. As the transmitter is placed further from the receiver, the contribution from the tail surpasses greatly the contribution from the peak, hence smoothing out the distinct trend shift for $r \simeq 2/3 l_c$. An evidence for this phenomenon can be observed from the relatively large values of average and half times compared to lower values of the peak times, as the difference, as depicted in Fig. \ref{fig:meantime}, is approximately an order of magnitude. 

Finally, we conclude by noting that by incorporating a reflecting boundary, one can describe the impulse response behavior of a cylindrical channel in a more realistic and exact manner. Through our analytical approach, we develop a formalism allowing comparisons between unbounded and bounded channels, as well as pawing the way for exploring the time-dependent response of the unbounded 2-D circular channel, where we can approximately widen our results to the unbounded case under the condition $D_0>> r_0$, $d_0$ and $D_0>> \sqrt{Dt}$. As a future work, we plan to explore the angular dependent impulse response of a point transmitter and the corresponding channel characteristics.

\section*{Acknowledgements}
Research at the Perimeter Institute is supported by the Government of Canada through the Department of Innovation, Science and Economic Development Canada, and by the Province of Ontario through the Ministry of Research and Innovation.

\appendix 

To describe the diffusion of the molecule inside the annular region under angle dependent conditions, we shall find a solution to the Fick's Law, satisfying the necessary boundary conditions
\begin{subequations}  
\begin{align}
\frac{\partial \Prob}{\partial r}\Big|_{r=D_0} &= 0,  \\ 
 \Prob\Big|_{r=d_0} &= 0,\\
 P( r,0|r_0)&=\frac{1}{ r} \delta(r-r_0) \delta(\theta- \theta_0),
\end{align}
\end{subequations}
where we recall that, since the boundaries are described by both Neumann and Dirichlet boundary conditions, the Laplacian operator ($\Laplace$) is guaranteed to have a unique solution. 

We shall start with the separation of variables ansatz 
\begin{equation}
\Prob=\phi(r,\theta) T(t),
\end{equation}
which leads to 
\begin{equation*}
T(t) = e^{-\mu^2 t}
\end{equation*}
and
\begin{align*}
\frac{R''}{R} + \frac{R'}{rR} + \frac{\Theta''}{r^2 \Theta} = - \frac{\mu^2}{D}.
\end{align*}
Let us now set:
\begin{align*}
\Theta'' = - m^2 \Theta \implies\Theta(\theta) = A_m \cos(m \theta) + B_m \sin(m\theta).
\end{align*}
Then, the solution to the radial equation becomes
\begin{equation}
R(r) = J_m\left(\frac{\mu}{\sqrt{D}} r\right) + c Y_m \left(\frac{\mu}{\sqrt{D}} r\right).
\end{equation}

Here, we shall define the function $\eta_m(\beta_{mn} x)$ as
\begin{equation}
\eta_m(\beta_{mn} x) = J_m (\beta_{mn} x) + c_{nm} Y_m(\beta_{mn} x)
\end{equation}
such that $\eta_m(\beta_{mn})' =0$ and $\eta_m(\beta_{mn} \alpha)=0$, where $\alpha=\frac{d_0}{D_0}$ as usual. Then, $\eta_m(\beta_{mn} \frac{r}{D_0})$ are indeed solutions to the radial equation with the boundary conditions satisfied, where $\beta_{mn} = \frac{D_0 \mu}{\sqrt{D}}$. In general, to find $\beta_{mn}$, we shall solve a linear set of equations similar to what we have done for $\eta_0(\beta_n x)$.

Before continuing, we shall give the normalization condition for the special function $\eta_m(\beta_{mn} x)$ as
\begin{align*}
&\int_\alpha^1 \eta_m(\beta_{mn} x) \eta_{m'} (\beta_{n'm'}x) x \diff x = I_{mn} \delta_{nn'} \delta_{mm'},
\end{align*}
where we note that $I_{mn}$ can be written in terms of linear combinations of Bessel functions of the first and second kind. Without loss of generality, we can set $\theta_0=0$ and find the probability density function for the molecules as

\scriptsize
\begin{equation}
    \begin{split}
        &\Prob = \sum_{ n=1}^\infty \frac{1}{2 \pi D_0^2 I_{0n}} \eta_0\left(\beta_{0n} \frac{r_0}{D_0} \right) \eta_0 \left(\beta_{0n} \frac{r}{D_0}\right) e^{-\beta_{0n}^2 \frac{Dt}{D_0^2}}+ \\
&\sum_{m=1, n=1}^\infty \frac{1}{\pi D_0^2 I_{mn}} \cos(m \theta) \eta_m\left(\beta_{mn} \frac{r_0}{D_0} \right) \eta_m \left(\beta_{mn} \frac{r}{D_0}\right) e^{-\beta_{mn}^2 \frac{Dt}{D_0^2}}. \label{eq:probfinal}
    \end{split}
\end{equation}
 \normalsize
Some $\beta_{mn}$ values are given in Table 1 in the following page.

\begin{table*}
\centering
\caption{$\beta_{mn}$ Values for $\alpha=0.1$ Calculated By Our Algorithm}
\label{my-label}
\scriptsize
\begin{tabular}{|l|l|l|l|l|l|l|l|l|l|l|l|l|l|}
\hline
    & n=1    & n=2    & n=3    & n=4    & n=5    & n=6    & n=7    & n=8    & n=9     & n=10     & n=11    & n=12    & n=13    \\ \hline
m=0  & 1.103  & 4.979  & 8.554  & 12.087 & 15.603 & 19.111 & 22.614 & 26.114 & 29.612  & 33.108  & 36.604  & 40.099  & 43.593  \\ \hline
m=1  & 1.879  & 5.532  & 8.975  & 12.422 & 15.880 & 19.346 & 22.818 & 26.293 & 29.772  & 33.253  & 36.736  & 40.219  & 43.704  \\ \hline
m=2  & 3.056  & 6.724  & 10.042 & 13.347 & 16.677 & 20.038 & 23.424 & 26.831 & 30.254  & 33.689  & 37.133  & 40.584  & 44.041  \\ \hline
m=3  & 4.201  & 8.016  & 11.353 & 14.612 & 17.858 & 21.118 & 24.402 & 27.714 & 31.054  & 34.416  & 37.798  & 41.195  & 44.606  \\ \hline
m=4  & 5.318  & 9.282  & 12.682 & 15.967 & 19.206 & 22.428 & 25.650 & 28.886 & 32.142  & 35.423  & 38.728  & 42.056  & 45.404  \\ \hline
m=5  & 6.416  & 10.520 & 13.987 & 17.313 & 20.576 & 23.807 & 27.020 & 30.227 & 33.437  & 36.657  & 39.894  & 43.151  & 46.430  \\ \hline
m=6  & 7.501  & 11.735 & 15.268 & 18.637 & 21.932 & 25.184 & 28.411 & 31.622 & 34.823  & 38.021  & 41.222  & 44.431  & 47.653  \\ \hline
m=7  & 8.578  & 12.932 & 16.529 & 19.942 & 23.268 & 26.545 & 29.791 & 33.016 & 36.226  & 39.426  & 42.620  & 45.812  & 49.006  \\ \hline
m=8  & 9.647  & 14.116 & 17.774 & 21.229 & 24.587 & 27.889 & 31.155 & 34.397 & 37.620  & 40.831  & 44.031  & 47.225  & 50.414  \\ \hline
m=9 & 10.711 & 15.287 & 19.005 & 22.501 & 25.891 & 29.219 & 32.505 & 35.764 & 39.002  & 42.225  & 45.436  & 48.637  & 51.832  \\ \hline
m=10 & 11.771 & 16.448 & 20.223 & 23.761 & 27.182 & 30.535 & 33.842 & 37.118 & 40.371  & 43.607  & 46.829  & 50.040  & 53.243  \\ \hline
m=11 & 12.826 & 17.600 & 21.431 & 25.009 & 28.461 & 31.838 & 35.167 & 38.460 & 41.729  & 44.978  & 48.211  & 51.433  & 54.645  \\ \hline
m=12 & 13.879 & 18.745 & 22.629 & 26.246 & 29.729 & 33.131 & 36.481 & 39.792 & 43.075  & 46.338  & 49.583  & 52.816  & 56.037  \\ \hline
m=13 & 14.928 & 19.883 & 23.819 & 27.474 & 30.987 & 34.415 & 37.784 & 41.114 & 44.412  & 47.688  & 50.946  & 54.189  & 57.420  \\ \hline
m=14 & 15.975 & 21.015 & 25.002 & 28.694 & 32.237 & 35.689 & 39.079 & 42.426 & 45.740  & 49.030  & 52.299  & 55.553  & 58.794  \\ \hline
m=15 & 17.020 & 22.142 & 26.178 & 29.907 & 33.478 & 36.954 & 40.365 & 43.730 & 47.059  & 50.363  & 53.644  & 56.909  & 60.160  \\ \hline
m=16 & 18.063 & 23.264 & 27.347 & 31.112 & 34.712 & 38.212 & 41.643 & 45.025 & 48.371  & 51.687  & 54.982  & 58.257  & 61.518  \\ \hline
m=17 & 19.104 & 24.382 & 28.511 & 32.311 & 35.940 & 39.463 & 42.914 & 46.314 & 49.674  & 53.005  & 56.311  & 59.598  & 62.869  \\ \hline
m=18 & 20.144 & 25.496 & 29.670 & 33.504 & 37.160 & 40.707 & 44.178 & 47.595 & 50.971  & 54.315  & 57.634  & 60.932  & 64.213  \\ \hline
m=19& 21.182 & 26.606 & 30.824 & 34.691 & 38.375 & 41.945 & 45.436 & 48.870 & 52.261  & 55.619  & 58.950  & 62.259  & 65.550  \\ \hline
m=20 & 22.219 & 27.712 & 31.974 & 35.874 & 39.585 & 43.177 & 46.687 & 50.139 & 53.545  & 56.916  & 60.260  & 63.580  & 66.881  \\ \hline
m=21 & 23.255 & 28.816 & 33.119 & 37.052 & 40.789 & 44.403 & 47.933 & 51.401 & 54.823  & 58.208  & 61.563  & 64.895  & 68.206  \\ \hline
m=22 & 24.289 & 29.916 & 34.261 & 38.225 & 41.988 & 45.624 & 49.173 & 52.659 & 56.095  & 59.494  & 62.861  & 66.204  & 69.525  \\ \hline
m=23 & 25.323 & 31.014 & 35.399 & 39.394 & 43.183 & 46.841 & 50.409 & 53.911 & 57.362  & 60.774  & 64.154  & 67.507  & 70.839  \\ \hline
m=24 & 26.356 & 32.109 & 36.533 & 40.559 & 44.373 & 48.053 & 51.639 & 55.158 & 58.624  & 62.049  & 65.441  & 68.806  & 72.148  \\ \hline
m=25 & 27.387 & 33.202 & 37.665 & 41.721 & 45.559 & 49.260 & 52.865 & 56.400 & 59.881  & 63.320  & 66.724  & 70.099  & 73.451  \\ \hline
m=26 & 28.418 & 34.293 & 38.793 & 42.879 & 46.742 & 50.463 & 54.087 & 57.638 & 61.134  & 64.585  & 68.001  & 71.388  & 74.750  \\ \hline
m=27 & 29.448 & 35.382 & 39.919 & 44.033 & 47.920 & 51.663 & 55.305 & 58.872 & 62.382  & 65.847  & 69.275  & 72.672  & 76.045  \\ \hline
m=28 & 30.478 & 36.468 & 41.042 & 45.185 & 49.096 & 52.859 & 56.518 & 60.101 & 63.626  & 67.104  & 70.543  & 73.952  & 77.334  \\ \hline
m=29 & 31.506 & 37.553 & 42.163 & 46.333 & 50.268 & 54.051 & 57.728 & 61.327 & 64.866  & 68.356  & 71.808  & 75.228  & 78.620  \\ \hline
m=30 & 32.534 & 38.636 & 43.281 & 47.479 & 51.436 & 55.239 & 58.934 & 62.549 & 66.102  & 69.605  & 73.069  & 76.499  & 79.902  \\ \hline
m=31 & 33.562 & 39.717 & 44.397 & 48.622 & 52.602 & 56.425 & 60.137 & 63.768 & 67.334  & 70.851  & 74.326  & 77.767  & 81.179  \\ \hline
m=32 & 34.588 & 40.797 & 45.510 & 49.762 & 53.765 & 57.607 & 61.337 & 64.982 & 68.563  & 72.092  & 75.579  & 79.031  & 82.453  \\ \hline
m=33 & 35.615 & 41.875 & 46.622 & 50.900 & 54.925 & 58.787 & 62.533 & 66.194 & 69.789  & 73.330  & 76.828  & 80.291  & 83.723  \\ \hline
m=34 & 36.641 & 42.952 & 47.731 & 52.036 & 56.083 & 59.963 & 63.727 & 67.403 & 71.011  & 74.565  & 78.075  & 81.548  & 84.990  \\ \hline
m=35 & 37.666 & 44.028 & 48.839 & 53.169 & 57.238 & 61.137 & 64.917 & 68.608 & 72.230  & 75.796  & 79.317  & 82.801  & 86.253  \\ \hline
m=36 & 38.691 & 45.102 & 49.945 & 54.301 & 58.390 & 62.308 & 66.105 & 69.811 & 73.446  & 77.024  & 80.557  & 84.051  & 87.513  \\ \hline
m=37 & 39.715 & 46.174 & 51.049 & 55.430 & 59.541 & 63.477 & 67.290 & 71.010 & 74.659  & 78.250  & 81.794  & 85.298  & 88.770  \\ \hline
m=38 & 40.739 & 47.246 & 52.152 & 56.557 & 60.689 & 64.643 & 68.472 & 72.207 & 75.869  & 79.472  & 83.027  & 86.542  & 90.023  \\ \hline
m=39 & 41.762 & 48.317 & 53.253 & 57.682 & 61.835 & 65.807 & 69.652 & 73.401 & 77.076  & 80.692  & 84.258  & 87.783  & 91.274  \\ \hline
m=40 & 42.785 & 49.386 & 54.352 & 58.806 & 62.978 & 66.969 & 70.829 & 74.593 & 78.281  & 81.908  & 85.486  & 89.021  & 92.521  \\ \hline
m=41 & 43.808 & 50.454 & 55.450 & 59.927 & 64.120 & 68.128 & 72.005 & 75.782 & 79.483  & 83.122  & 86.711  & 90.257  & 93.766  \\ \hline
m=42 & 44.830 & 51.521 & 56.546 & 61.047 & 65.260 & 69.285 & 73.177 & 76.969 & 80.683  & 84.334  & 87.933  & 91.489  & 95.008  \\ \hline
m=43 & 45.852 & 52.588 & 57.642 & 62.166 & 66.398 & 70.441 & 74.348 & 78.154 & 81.880  & 85.543  & 89.153  & 92.719  & 96.248  \\ \hline
m=44 & 46.874 & 53.653 & 58.735 & 63.282 & 67.534 & 71.594 & 75.517 & 79.336 & 83.075  & 86.750  & 90.371  & 93.947  & 97.485  \\ \hline
m=45 & 47.895 & 54.717 & 59.828 & 64.397 & 68.669 & 72.745 & 76.683 & 80.517 & 84.268  & 87.954  & 91.586  & 95.172  & 98.719  \\ \hline
m=46 & 48.916 & 55.781 & 60.919 & 65.511 & 69.801 & 73.895 & 77.848 & 81.695 & 85.459  & 89.156  & 92.798  & 96.394  & 99.951  \\ \hline
m=47 & 49.937 & 56.843 & 62.009 & 66.623 & 70.932 & 75.042 & 79.010 & 82.871 & 86.647  & 90.356  & 94.009  & 97.614  & 101.180 \\ \hline
m=48 & 50.958 & 57.905 & 63.098 & 67.734 & 72.062 & 76.188 & 80.171 & 84.045 & 87.834  & 91.553  & 95.217  & 98.832  & 102.407 \\ \hline
m=49 & 51.978 & 58.966 & 64.186 & 68.844 & 73.190 & 77.333 & 81.330 & 85.217 & 89.018  & 92.749  & 96.423  & 100.048 & 103.632 \\ \hline
m=50 & 52.998 & 60.026 & 65.273 & 69.952 & 74.316 & 78.475 & 82.487 & 86.387 & 90.200  & 93.943  & 97.627  & 101.262 & 104.855 \\ \hline
m=51 & 54.017 & 61.086 & 66.358 & 71.059 & 75.441 & 79.616 & 83.642 & 87.556 & 91.381  & 95.134  & 98.828  & 102.473 & 106.076 \\ \hline
m=52 & 55.037 & 62.145 & 67.443 & 72.164 & 76.565 & 80.756 & 84.796 & 88.722 & 92.559  & 96.324  & 100.028 & 103.683 & 107.294 \\ \hline
m=53 & 56.056 & 63.203 & 68.527 & 73.269 & 77.687 & 81.894 & 85.948 & 89.887 & 93.736  & 97.511  & 101.226 & 104.890 & 108.510 \\ \hline
m=54 & 57.075 & 64.260 & 69.609 & 74.372 & 78.808 & 83.030 & 87.099 & 91.051 & 94.911  & 98.697  & 102.422 & 106.096 & 109.725 \\ \hline
m=55 & 58.093 & 65.317 & 70.691 & 75.474 & 79.928 & 84.165 & 88.248 & 92.212 & 96.084  & 99.881  & 103.616 & 107.299 & 110.937 \\ \hline
m=56 & 59.112 & 66.373 & 71.772 & 76.575 & 81.046 & 85.299 & 89.395 & 93.372 & 97.256  & 101.064 & 104.809 & 108.501 & 112.148 \\ \hline
m=57 & 60.130 & 67.428 & 72.852 & 77.675 & 82.163 & 86.431 & 90.541 & 94.531 & 98.426  & 102.244 & 105.999 & 109.701 & 113.356 \\ \hline
m=58 & 61.148 & 68.483 & 73.931 & 78.774 & 83.279 & 87.562 & 91.686 & 95.688 & 99.594  & 103.423 & 107.188 & 110.899 & 114.563 \\ \hline
m=59 & 62.166 & 69.537 & 75.010 & 79.872 & 84.394 & 88.692 & 92.829 & 96.843 & 100.761 & 104.601 & 108.375 & 112.095 & 115.768 \\ \hline
\end{tabular}
\end{table*}

\end{document}